\documentclass[%
 reprint,
groupedaddress,
showpacs,
 amsmath,amssymb,
 aps,
prb,
]{revtex4-1}

\usepackage{graphicx}
\usepackage{dcolumn}
\usepackage{bm}

\begin{document}

\newcommand{\be}{\begin{equation}}
\newcommand{\ee}[1]{\label{#1}\end{equation}}
\newcommand{\bem}{\begin{eqnarray}}
\newcommand{\eem}[1]{\label{#1}\end{eqnarray}}
\newcommand{\eq}[1]{Eq.~(\ref{#1})}
\newcommand{\Eq}[1]{Equation~(\ref{#1})}
\newcommand{\vp}[2]{[\mathbf{#1} \times \mathbf{#2}]}


\title{Symmetry of Kelvin-wave dynamics and the Kelvin-wave cascade in the $T=0$ superfluid turbulence}

\author{E.  B. Sonin}
 \affiliation{Racah Institute of Physics, Hebrew University of
Jerusalem, Givat Ram, Jerusalem 91904, Israel}

\date{\today}

\begin{abstract}
The article considers implications of tilt symmetry (symmetry with respect to tilting of the coordinate  axis with respect to which  vortex motion is studied)  in the non-linear dynamics of Kelvin waves. The conclusion is that although the spectrum of Kelvin wave is not tilt-invariant, this does not compromise tilt invariance of the Kelvin-wave cascade vividly argued now in the theory of superfluid turbulence. The article investigates the effect of strong kelvon interaction on the power-law exponent for the Kelvin-wave cascade and suggests a simple picture of the crossover from the classical Kolmogorov cascade to the quantum Kelvin-wave cascade, which does not encounter with mismatch of the energy distributions at the crossover and does not require a broad intermediate interval for realization of the crossover.
\end{abstract}

\maketitle


\section{Introduction}

During decades superfluid turbulence remains in the focus of theoretical and experimental investigations of various superfluids, in particular,  $^4$He, $^3$He, and Bose-Einstein condensates.\cite{Turb,Naz}   Special attention was drawn to superfluid turbulence at very low temperatures, where damping of turbulence by mutual friction is expected to be very small.   Vinen's theory of superfluid turbulence  characterizes  the turbulent vortex tangle by the   {\em length per unit volume} $\cal L$ with dimensionality cm$^{-2}$.  The length determines  the  distance between lines and their curvature radius both estimated by the scale $l_0 \sim {\cal L}^{-1/2}$. Despite the crucial role of quantum  circulation in superfluid turbulence, one may expect that it will not have any impact on turbulence at spatial scales longer than the typical intervortex distance $l_0$ of the vortex tangle.
Then superfluid turbulence should not differ from classical superfluid and must be described by the classical Kolmogorov cascade \cite{VN}. This was confirmed by a number of numerical and physical experiments starting from those done more than a decade ago \cite{NAB,MT}. The concept of the cascade is based on the assumption that energy pumped into a turbulent liquid at large length scales is transferred to very small scales without losses by means of the constant energy flux $\varepsilon$ in the space of inverse scales (wave numbers $k$). 

But the classical turbulence cannot extend to scales $l$ shorter than the intervortex spacing $l_0$ in the vortex tangle. 
At scales  $l \ll l_0$ it is natural to expect that the main mechanism of the energy relaxation (transfer to  smaller scales of the vortex tangle) is Kelvin modes related with distortions of vortex lines.  The energy flux transfers the energy  from lower to higher Kelvin-wave numbers without dissipation (pure Kelvin-wave cascade) up to very high wave numbers, at which the energy ultimately dissipates via phonon radiation.\cite{Vin00,Kiv} Following this scenario, 
 Kozik and Svistunov \cite{Koz} derived the Kelvin-wave cascade from the Boltzmann equation taking into account scattering of three kelvons (six-wave interaction). The processes involving less kelvons are forbidden by  the conservation laws of  energy and momentum.
However, \citet{Lau} (see also \citet{Lvo}) challenged this theory arguing that  relevant integrals determining the energy flux diverge, if one of $k$, which determine  vertices in the Boltzmann equation, goes to zero and vertices depend on this small $k$ {\em linearly}. 
This means failure of the central assumption of the cascade concept: {\em locality}.  They suggested a modified cascade scenario treating  a long-wavelength mode as a quasistatic field, which allows four-wave interaction with large $k$, since the long-wavelength mode  lifts restrictions imposed by the conservation laws.  They concluded that despite failure of the locality assumption the spectrum in the new cascade is universal. 
Kozik and Svistunov \cite{Koz10} pointed out that  symmetry with respect to rotations of the vortex line around the axis normal to the line (tilt symmetry) rules out linear dependence  of any physically meaning quantity on a mode with very small $k$, because the latter is equivalent to tilting of the coordinate axis.  So the correct  asymptote  at small $k$ must be not $k$ but $k^2$, which corresponds to dependence on the vortex-line curvature (instead of the dependence on a tilt angle). This controversy led  to vivid discussion \cite{Leb,KozC,LLN,Lvo11}. Lebedev {\em et al.} \cite{Leb,LLN} and Bou\'{e} {\em et al.} \cite{Lvo11} insisted that linear dependence on $k$ is not forbidden by symmetry, since  tilt symmetry is broken {\em globally}, i.e. by boundary conditions at very long scales. 

The difference between the cascade power-law exponents  predicted by the two approaches was not so big (power-law exponent 3.4 vs. 3.67 for distribution of kelvons in the $k$ space), especially keeping in  mind difficulties in accurate extraction of the exponent from numerical and physical experiments. But it is the case when the source of disagreement is much more important than disagreement itself. The dispute is about the fundamental of non-linear vortex dynamics: what is proper symmetry and how it manifests itself.  Moreover, the absence of locality must lead to dramatic consequences for non-linear Kelvin wave dynamics.

 In general, one cannot exclude a chance that  in the absence of locality  broken symmetry at long scales breaks also symmetry at short scales, and the boundary conditions affect symmetry of  an infinitely long system in the thermodynamic limit as L'vov {\em et al.} believe. However, there is a  logical vicious circle here. By believing that global symmetry wins out over local symmetry (i.e., symmetry of the hamiltonian), L'vov {\em et al.} assume the absence of locality, while locality can be absent only if local tilt symmetry is overcome by lower global symmetry. The way out from the circle is first to forget about possible effect of global symmetry.  The latter can enter into the game only if local symmetry cannot guarantee locality.  This justifies  reassessment of implications of {\em local} tilt symmetry for the problem of non-linear Kelvin waves undertaken in this paper.  The analysis  focuses on some aspects, which were not addressed in former debates. \cite{Koz10,Leb,KozC,LLN} 

The paper  analyzes  the effect of truncation of the hamiltonian on its tilt symmetry.   Although truncation always breaks tilt symmetry, after a proper truncation  symmetry is broken only in terms beyond accuracy of the truncation. It is argued that truncation undertaken by  L'vov {\em et al.} is not of this kind. Therefore, their challenge of the locality assumption is not reliable, being in a conflict with local symmetry.  
Further, the  paper considers the Kelvin wave of arbitrary amplitude, which has an exact solution  in the local-induction approximation, and therefore the problem of truncation and renormalization of expansion series is eliminated. Nevertheless, the  spectrum of the non-linear Kelvin wave depends on the tilt of the vortex line towards an axis, with respect to which displacements of the vortex line are measured. One could consider this as vindicating the view of L'vov {\em et al.} who noticed evidence of this  effect in the perturbation-theory expansion \cite{LLN}. However, the effect reflects the  trivial geometrical fact that the Kelvin wavelength along the vortex line does not coincide with its projection on an axis arbitrarily  tilted to the unperturbed vortex line. In a sense it is an analogue of the Doppler effect: the wave spectrum depends on the coordinate frame in which it is observed, despite physical equivalence of all frames. But a real physical process cannot appear in one frame and disappear in another. The mechanism of  L'vov {\em et al.} is absent in the coordinate  frame with the axis coinciding with the average position of the vortex line, in which average vortex displacement and tilt are absent. This would be a mystery if the mechanism revived in any other frame.

Both competing approaches assumed weak non-linearity and considered kelvon interactions of the lowest possible order. There were a number of numerical simulations for the Kelvin-wave cascade but they  could not resolve the controversy  supporting either this or that stance or being in disagreement with the both. The paper suggests that a variety of exponents revealed in simulations is because the assumption of weak non-linearity is not valid everywhere in the Kelvin-wave cascade interval, and the value of the exponent can depend on the energy input into the cascade.  
The scaling-based derivation of the exponent \cite{Koz} was generalized on the case of $n$-kelvon interaction, which yielded an interval of possible exponents. Most of  exponents obtained in numerical simulations lie in this interval.

The last topic of the present paper is the crossover from the classical Kolmogorov cascade to the quantum Kelvin-wave cascade, which was also vividly argued in the literature. The paper revises the conditions for the crossover and suggests a scenario using only  a single scale:  tangle  scale $l_0$. This scenario is simpler than those discussed earlier, does not encounter with the problem of the mismatch of energy distributions at the crossover, and does not require a broad intermediate interval between the two cascades dependent on parameters other than $l_0$.

\section{Tilt-symmetry and truncation of the series expansion of the Kelvin-wave hamiltonian}\label{trunk}

In an attempt to demonstrate primacy of global symmetry over local symmetry,   \citet{Leb} considered the vortex line length, which is a hamiltonian for the vortex line  in the local-induction approximation.  Here we revise the analysis by \citet{Leb} arriving at new  conclusions. Even though a cascade  in the local-induction  approximation is impossible,  studying of symmetry in this model is much simpler but still relevant for   
the Bio--Savart dynamics allowing the Kelvin-wave cascade.    

The vortex-line length is given by the functional
\be
L=\int \sqrt{1+\left(\frac{\partial \bm u }{\partial z}\right)^2 }dz, 
   \ee{L}
where $\bm u(x,y)$ is a two-dimensional displacement vector in the $xy$ plane. In  the local-induction approximation the canonical Hamiltonian equations of motion are
\be
[\hat z \times \bm u]=-\nu_s {\delta L \over \delta  \bm u}=\nu_s  {\partial \over \partial z}\left[\frac{\partial \bm u /\partial z}{ \sqrt{1+\left(\partial \bm u /\partial z\right)^2 }}\right],
   \ee{EM}
where $\nu_s =\kappa\Lambda/4\pi$ is the line-tension parameter,  $\kappa$ is the circulation quantum, $\Lambda = \ln (r_m/r_c)$ is a large logarithmic factor, $r_c$ is the vortex core radius, and $r_m$ is the scale cutting off the velocity field $\sim 1/r$ at large $r$ (the curvature radius, or the intervortex distance). The functional (\ref{L}) and the equations of motion (\ref{EM}) are invariant with respect to a tilt of the vortex axis resulting from rotation around the axis $y$ through the angle $ \phi$:
\bem
 x= \tilde x \cos \phi - \tilde z \sin \phi ,
\nonumber \\
 z= \tilde  x \sin \phi+ \tilde z \cos \phi. 
   \eem{rot}
But any truncation of the series expansion of the functional,  such as
\bem
L =\int \left\{1+{1\over 2}\left(\frac{\partial \bm u }{\partial z}\right)^2-{1\over 8}\left(\frac{\partial \bm u }{\partial z}\right)^4...\right\} dz ,
   \eem{exp}
does not respect  tilt symmetry.  
Let us write down the truncated functional \eq{exp} after rotation  keeping only terms linear in the rotation angle $\phi$    and assuming for the sake of simplicity that the vortex line lies in the $xz$ plane ($y=0$):
\bem
L =\int \left\{
\left[1-\phi {d\tilde x\over d\tilde z}  \right] 
\right.  \nonumber \\ \left.
 + {1\over 2}\left[ \left( {d\tilde x \over d\tilde z}\right)^2  + 2 \phi {d\tilde x \over d\tilde z}+ \phi\left({d\tilde x \over d\tilde z} \right)^3 \right]
\right.  \nonumber \\ \left.
 -{1\over 8}\int d\tilde z\left[ \left( {d\tilde x \over d\tilde z}\right)^4  + 4 \phi \left( {d\tilde x \over d\tilde z}\right)^3+3 \phi\left({d\tilde x \over d\tilde z} \right)^5 \right]
\right\}
d\tilde z. 
   \eem{expR}
The terms in square brackets show the zero-order, the quadratic and the quartic terms of \eq{exp} after rotation, which includes also transformation of the integral variable: $dz = [1-\phi (d\tilde x/ d\tilde z) ]d\tilde z$.  Any such term is not tilt-invariant, but if 
one keeps terms up to the quadratic one, only the higher-order  term $\propto \phi (d\tilde x/ d\tilde z)^3$ breaks  tilt symmetry, whereas keeping terms up to quartic, symmetry is broken by the higher-order term  $\propto \phi (d\tilde x/ d\tilde z)^5$.  Thus, although  truncation of the tilt-invariant hamiltonian breaks tilt invariance, symmetry is broken only by higher-order terms, which are beyond accuracy of the approximation.  Estimating the order of terms, one should take into account that the tilt angle may be considered as the Kelvin mode in the limit of zero wave number $k$: $\phi \sim k a$, where $a$ is the amplitude of displacement produced by the Kelvin wave [see \eq{sol} below]. Thus the terms $\propto \phi (d\tilde x/ d\tilde z)^3$ and $\propto \phi (d\tilde x/ d\tilde z)^5$ breaking tilt symmetry are of the 4th and the 6th order in the Kelvin mode amplitude, respectively. 

This provides an indication that the model of  L'vov {\em et al.} violates tilt symmetry. In their analysis they used the so-called {\em local nonlinear equation} corresponding to the hamiltonian
\bem
H_{LNE} \propto \int \left\{1+{1\over 2}\left(\frac{\partial \tilde {\bm u} }{\partial z}\right)^2-{1\over 24}\left(\frac{\partial \tilde {\bm u}  }{\partial z}\right)^6...\right\} dz,
   \eem{lne}
which looks to be not tilt-invariant. However, the displacement $ \tilde {\bm u}$ is not a true displacement $\bm u$, but is obtained from the latter by some nonlinear transformation, and therefore the rule of its transformation at axis rotation is not self-evident. On the other hand, \citet{Lau} pointed out that this hamiltonian is isomorphic to the hamiltonian in the  truncated local-induction approximation given (up to a constant factor) by \eq{exp}. We have shown that this hamiltonian contains the 6th order terms violating tilt symmetry. Since the main contribution to Kelvin-wave dynamics comes from the 6th order terms, the hamiltonian used by  L'vov {\em et al.} violates tilt symmetry.

\section{Tilt-symmetry without truncation and broken tilt-symmetry of the spectrum of the Kelvin wave}

Because of problems with tilt symmetry of truncated hamiltonians, it is useful to check tilt symmetry for  a Kelvin wave of arbitrary amplitude, which has an exact solution without any truncation of the original tilt-invariant hamiltonian. A monochromatic Kelvin wave propagating along the axis $z$ in the original coordinate frame is   
\be
x=a \cos (kz -\omega t),~~ y=a \sin (kz -\omega t) .
     \ee{sol}
It exactly satisfies the equations of motion (\ref{EM}),  if the frequency is given by 
\be 
\omega=\frac{\nu_s k^2 }{\sqrt{1+a^2k^2}}.
     \ee{}
A vortex line with a high-amplitude Kelvin wave is a helical vortex, which was actively investigated in classical and superfluid hydrodynamics \cite{Saf,Hel}.

Now let us transform the Kelvin wave in \eq{sol} to the tilted coordinate frame:
\bem
\Delta x ={a\over \cos \phi}\cos\left({k \tilde z \over \cos \phi} -\omega t -k \Delta x \sin \phi\right)  ,
\nonumber \\
y=a \sin \left({k \tilde z \over \cos \phi} -\omega t -k \Delta x \sin \phi\right) .
  \eem{fr}
Here $\Delta x =\tilde x+\tilde z \tan\phi$ is a displacement  produced by the Kelvin wave with respect to the line $\tilde x=-\tilde z \tan\phi$ (a vortex line undisturbed by the Kelvin wave).

\begin{figure}[t]
 \includegraphics[width=0.75\linewidth]{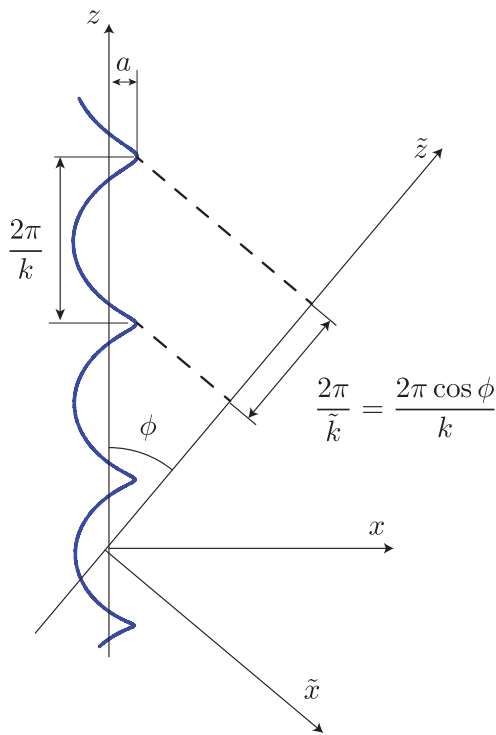}%
 \caption{Kelvin wave in the coordinate frame with the $z$ axis along an unperturbed straight vortex line and  in the rotated coordinate frame with the $\tilde z$ axis tilted to the original  $z$ axis by the angle $\phi$. The spatial periods  $2\pi/k$ and $2\pi/\tilde k$ are different in the two frames.}  \label{f1}
 \end{figure}

Even a weak Kelvin wave, when one may neglect the term $\propto \Delta x$ in arguments of trigonometric functions, is essentially different from that in the original coordinate frame: the wave is elliptically but not circularly polarized, and the wave number $k$ transforms to $k/\cos\phi$. So the value of $k$ depends on the choice of the coordinate system. This dependence directly follows from the geometry of the problem: the period $2\pi /k$ of the wave along the vertical axis in the original coordinate frame is by a factor $1/\cos \phi$ longer than the period along the tilted axis (Fig.~\ref{f1}). 
If the wave is non-linear and  the term $\propto \Delta x$ in arguments becomes important, the wave is not harmonic anymore and subharmonics appear in the $k$ space. In summary, tilting of the axis affects distribution in the $k$ space.

\citet{LLN} revealed evidence of this effect in the series expansion in the $k$ space and called it the nonlinear shift of the Kelvin wave frequency with the wave of small $k$. They  argued that this was a  nontrivial  {\em observable physical} effect, which  supported their stance.  Without arguing observability of the effect,  I would prefer to call it a {\em visual} rather than physical effect, which has nothing to do with the global symmetry at the border. It originates from not the most optimal choice of the coordinate axis. In reality nothing happened with the original Kelvin wave {\em physically} and its frequency was not shifted, but the wavelength of  the mode was measured by a different spatial scale. One can directly check with the equation of motion that the relation (\ref{fr}) for  a frequency for a nonlinear wave is valid after transformation to the tilted coordinate frame after scaling of $k$, i.e., replacing it with $\tilde k =k/\cos\phi$. 

Scaling of $k$ does not affect the Kelvin-wave exponent,  the latter being the same in the $k$ and the $\tilde k$ space. The mechanism of L'vov {\em et al.} originates  from  the quasistatic Kelvin mode, which in the limit of small $k$ equivalent to a tilt $\phi \sim ka$ of the $z$ axis. The tilt can be removed by  transformation to another coordinate frame, in which the mechanism  disappears. It is impossible to imagine how a real physical mechanism  of kelvon interaction resurrected simply by tilting the coordinate axis. Unfortunately it is very difficult to check this by explicit calculations. \citet{Lvo11} reported that they took into account 72(!) terms of the sixth order in their calculation. If one  also considered other 6th-order contributions as discussed in the previous section, this number would be essentially more. But this is exactly a situation in which benefits of the symmetry approach can be exploited. If symmetry requires that some contribution must vanish, one may believe this even without lengthy calculation. 

 The dependence of the Kelvin-wave spectrum on the tilt angle between the vortex line and an arbitrary chosen coordinate  axis is similar to the Doppler effect. But an essential difference between them is that in the Doppler effect transition to another coordinate system changes frequency without affecting wavelength, while tilting of the coordinate axis changes wavelength without affecting frequency.

\section{The effect of strong kelvon interaction on the exponent of the Kelvin-wave cascade}

The assumption of locality being not discredited, it is worthwhile to address another assumption also used at derivation of the Kelvin-wave cascade. Calculating the energy flux in the 1D $k$ space,  \citet{Koz08}  (as well as L'vov {\em et al.}), assumed that the non-linearity of the waves is weak and the first non-trivial collision term in the perturbation theory  (six-waves interaction) determines the dynamics of the Kelvin-wave cascade.  \citet{Koz08,Koz09}  stressed that it is asymptotically exact in the limit of high wave numbers, or inverse length scales. However, the interval of the  Kelvin-wave cascade in the $k$ space extends to lower wave numbers where the parameter of the perturbation theory is not small, and one may expect that the higher-order terms can be important as well, or even more important.  In order to check possible consequences of this, one may generalize the scaling estimation  of \citet{Koz}  for  for 3-kelvon interaction on the case when $n$ kelvons participate in the collision. 

Let us introduce the intensity of the Kelvin mode $m(k)=|\bm u(k)|^2 $, where $ \bm u(k)  $ is the Fourier component of the  displacement expansion $\bm u(z) =\tilde L^{-1/2} \sum\limits_k \bm u(k) e^{ik z} $ for the vortex line of length $\tilde L$. The intensity $m(k)$ is proportional to the occupation number  of kelvons and has dimensionality cm$^3$.
 The structure of the term in  the Boltzmann equation for  variation of $m(k)$ in time taking into account the $n$-kelvon interaction is 
\be 
\dot m(k) = \sum_{\{k_i\}} V (k_1,k_2,...k_{2n}) m(k_1) m(k_2)... m(k_{2n-1}),  
\ee{}
where summation over $\{k_i\}$ is summation over $2n-1$ wave numbers not equal to $k$, while one  argument for the   vertex $V (k_1,k_2,...k_{2n})$ must coincide with $k$. The vertex describes the interaction between $n$ Kelvin modes and contains $\delta$-functions $\delta(k_1+k_2+...+k_{2n})$ and $\delta(\omega_1+\omega_2+...+\omega_{2n})$ providing the conservation laws of energy and momentum \citep{Koz}.
The assumption of locality requires that all relevant  $k$ are of the same order, and a proper dimensionality is provided  if 
\be 
\dot m(k)  \sim  {\kappa  k ^ {6n-4} \over  \Lambda}m(k)^{2n-1} .
\ee{}
The large logarithm $\Lambda $ appears from scaling of $\delta(\sum \omega)$ by $1/\omega \sim 1/\kappa \Lambda k^2$. This scaling estimation allows to estimate the energy flux in the space of 1D wave numbers:
\bem 
{dE\over dt} = -\varepsilon= {\rho \kappa^2 \Lambda \tilde L \over 4\pi} \int^k dk'\, k'^2   \dot m(k') 
\nonumber \\
 \sim  {\rho \kappa^3\tilde  L \over 4\pi}  m(k)^{2n-1} k ^ {6n-1}, 
    \eem{KScas}
where $\rho$ is the 3D mass density. The cascade scenario assumes that the energy flux does not depend on $k$. The energy is calculated per unit volume, and the length $\tilde L$ is on the order of ${\cal L}\sim 1/l_0^2$. Eventually one obtains that
\be
 m(k) \sim \left(\varepsilon  l_0^2\over \rho \kappa^3\right)^{1\over 2n-1} k^{-{6n-1\over 2n-1}}.
   \ee{kel}
  For $n=2$ one obtains the distribution $ k^{-11/3 }$ of L'vov {\em et al.} while $n=3$ corresponds to  the distribution $ k^{-17/5}= k^{-3.4}$ of Kozik and Svistunov. If non-linearity is strong the case of  large $n$ is expected  and the distribution approaches to the law $ k^{-3 }$. Earlier the law $ k^{-3 }$ was obtained from different but also dimensional arguments \cite{VTM,Naz}.

So one should not expect an universal power law in the Kelvin-wave cascade, which may depend on power input into the cascade. 
Putting aside the scenario by L'vov {\em al.} as problematic from the symmetry point of view, the power laws between $ k^{-3.4}$ and $ k^{-3 }$ are possible. The latter limit was apparently realized in computer simulation by \citet{VTM}, while numerical simulations by \citet{BB} revealed the power laws from  $ k^{-3.21 }$ to $ k^{-3.1 }$. On the other hand, \citet{Koz05,Koz11} numerically confirmed the power law $k^{-3.4}$. At the same time, Fig.~2 of \citet{Koz05} shows smooth transformation of the  law $k^{-3.4}$ at high $k$ to  the  law $k^{-3}$ at low $k$. But the authors related the area of  the  law $k^{-3}$ with  initial conditions of their simulation and considered it as a transient law, which must eventually transform at   longer times to the law $k^{-3.4}$. Meanwhile, the present analysis suggests that the  law $k^{-3}$ is not transient but would  persist at longer times  since at low $k$ the Kelvin-wave amplitudes grow and become non-linear. Additional simulations are needed in order to resolve, which interpretation is true.

All these laws are within the expected interval. In the light of our discussion of symmetry it is important  that in numerical simulations of the Kelvin-wave cascade one used the equations of motion respecting tilting symmetry. The Bio--Savart law satisfies this condition. On the other hand,  \citet{Lvo11} received their power law $ k^{-11/3 }= k^{-3.67 }$ using the truncated Hamiltonian violating tilting symmetry (see Sec.~\ref{trunk}). Naturally one should not expect that  their simulation results respect this symmetry. 

\section{Crossover between the  Kolmogorov cascade and the Kelvin-wave cascade}

An interesting question, which was also vividly discussed, is how a crossover between the classic Kolmogorov cascade and the quantum Kelvin-wave cascade may occur. The crossover is expected at scales around $l_0$. 
Let us compare the energy densities of the Kolmogorov cascade and the Kelvin-wave cascade extrapolated to the crossover scale $l_0$. For the Kolmogorov cascade the energy density in the  $k$ space is 
\be
 e(k) \sim \rho \left(\varepsilon\over \rho\right) ^{2/3}  k ^{-5/3}, 
     \ee{KC}
where
\be 
\varepsilon=-{dE\over dt}  \sim{\rho v (l)^3 \over l} 
 \ee{EnFl}
is the energy flux  in the  $k$ space, and  $v(l)$ is the velocity obtained from averaging over the scales smaller than  $l\sim 1/k$. A proper energy density in the Kelvin-wave cascade is the kinetic-energy density $\rho \kappa ^2{\cal L}  /k=\rho \kappa ^2 /k l_0^2$  in the $k$ space related with the superfluid velocity induced by the vortex tangle of the length $\cal L$. This is the energy distribution for a straight vortex line, which retains  also for a vortex tangle as demonstrated analytically and numerically.\cite{Ara,Nem}
 Its extrapolation to $k\sim 1/l_0$ yields $ \rho\kappa^2 /l_0$, which is of the same order as the Kolmogorov-cascade energy density extrapolated to $k\sim 1/l_0$ from smaller $k$ if the energy flux is 
\be
\varepsilon = { \rho\kappa ^3\over l_0^4}.
  \ee{geps}
The same energy flux provides continuity of the vorticity at the crossover as was pointed out by \citet{LNR}.

It is necessary to check, up to what long scales can one use the theory of the Kelvin wave cascade.
For addressing this issue,  let us estimate total elongation $\Delta {\cal L}$ of the tangled vortex line  in the Kelvin-wave cascade.\cite{Vin00}    It is given by the integral 
\be
\Delta {\cal L} \sim {\cal L}  \int _{1/l_0}^\infty  k^2 m(k) dk  \sim  {\cal L}   \left({   \varepsilon l_0^4 \over    \kappa^3  \rho } \right)^{1\over 2n-1} .
\ee{spec6}
Here  \eq{kel}    was used. 
In the Kelvin-wave cascade the elongation $\Delta {\cal L}$ must not exceed the  length $\cal L$ of the tangle. Otherwise  displacements induced by Kelvin modes  are larger that the intervortex distance $l_0$. This signalizes a leading role of reconnections, which invalidate the conditions for the pure Kelvin-wave cascade.  
\Eq{spec6} shows that  the condition $\Delta {\cal L}\sim \cal L$ is reached at the scale $l_0$ if  the energy flux  is given by \eq{geps} independently from the number $n$ of kelvons in collisions.   Thus the   Kelvin-wave cascade theory can be used  up to the scale $l_0$, and one has a coherent picture of the crossover governed by the single scale $l_0$.

In the past  more complicated scenarios of the crossover were suggested.
\citet{LNR,Lvo8} argued that the crossover is impeded by mismatch of the energy distributions on the two sides of the crossover, and developed the theory of the crossover taking  the bottleneck into account.
However, the bottleneck problem  arose because \citet{LNR,Lvo8} tried to match 
 the Kolmogorov-cascade  energy density \eq{KC} 
with 
  the energy density  in the Kelvin-wave cascade in the 1D $k$ space. The latter is determined
  by the Kelvin-wave distribution $m(k)$ given by \eq{kel} and is equal to 
\be
e_{kw} \sim \Lambda\rho\kappa^2{\cal L} k^2 m(k) \sim \Lambda{\rho\kappa^2\over l_0^2} \left(\varepsilon  l_0^2\over \rho \kappa^3\right)^{1\over 2n-1} k^{-{2n+1\over 2n-1}}.
   \ee{}
At $k \sim 1/l_0$ and $\varepsilon$ given by  \eq{geps}, this yields  $e_{kw} \sim  \Lambda\rho\kappa^2  /l_0$, which by a large logarithm factor $\Lambda$ exceeds the Kolmogorov-cascade energy density extrapolated to the same scale.
However, these two energy distributions must not match each other because they are determined in {\em different} $k$  spaces.\cite{Kiv} The energy density $e_{kw}$ includes also the energy of very high Fourier components of the 3D velocity field induced by the vortex tangle at very short scales close to the core radius $r_c$. That is why $e_{kw}$  is proportional to  $\Lambda$ and has  no counterpart in the theory of the Kolmogorov cascade, which ignores singularity of the liquid velocity at vortex lines and uses the energy distribution over the modulus $k$ of the 3D wave vector $\bm k$ related with the 3D velocity field. The counterpart of the Kolmogorov energy in the Kelvin-wave cascade is the kinetic energy of the velocity induced by the vortex tangle, which is also determined in the 3D $k$ space. As demonstrated above, there is no mismatch  between these two counterparts.

Another picture of the crossover was suggested by \citet{Koz08}, who believed  that  the Kolmogorov cascade becomes invalid at the scale $l_c \sim l_0\Lambda^{1/2}$ essentially longer than $l_0$. So there is an intermediate range of scales much longer than $l_0$, where neither the Kolmogorov, nor the Kelvin-wave cascade is valid.  \citet{Koz08} determined the scale $l_c$ from the condition that  within the interval $l>l_c$ of the Kolmogorov cascade the liquid  velocity  should exceed the vortex velocity $\kappa \Lambda /l$ induced by the vortex line with curvature radius $\sim l$. The condition puts in two questions. First, it ignores that the velocity induced by a curved vortex line segment of large curvature radius $l$  can be much lower than $\kappa \Lambda /l$ (or even be of opposite sign) because of Kelvin modes excited at scales much smaller than $l$. This effect was revealed analytically and numerically for a helical-vortex ring, i.e., for a ring with a strong Kelvin wave around the ring.\cite{KM,BHT,Hel} 
Second, the condition is imposed on the velocity of a singular vortex line, and it is not clear whether  its violation 
can affect the energy  distribution in the  $k$ space of the continuous velocity field, which the theory of the Kolmogorov cascade addresses.

Summarizing, the previously suggested scenarios  of the crossover, being more complicated, do not look more reliable than our scenario  governed self-consistently only by the single scale $l_0$. Additional scales in the previous scenarios appeared due to large logarithm factor $\Lambda$, which results from the ultraviolet divergence at high $k$ in the velocity field induced by the vortex line. However, following the spirit of locality assumption, which excludes the effect of low $k$ on the cascade, it is not evident why very high $k$ must affect  the processes in the cascade either.

\section{Conclusions}

In summary, the conclusion of the present analysis is that although the spectrum of Kelvin waves {\em does depend} on the choice of the coordinate frame in which the Kelvin wave is investigated (the effect similar to the Doppler effect in moving media), the mechanism of the Kelvin-wave cascade must be tilt-invariant, i.e., invariant with respect to rotation of the axis along which the unperturbed vortex line presumably located. Therefore, the mechanism, which violates this invariance, is not credible and apparently does not  collect all symmetry-breaking terms in the series expansion, which eventually must cancel. The paper attracts attention to the fact that the assumption of weak non-linearity used in previous theoretical studies is not always valid and estimates how it can change the values of the cascade exponent.
Finally, the paper suggests a simple scenario of  the crossover between the classical Kolmogorov and the quantum Kelvin-wave cascades, which does not encounter with any problem of  mismatch or bottleneck at the crossover and does not require introduction of a broad intermediate interval between the cascades.

\begin{acknowledgments}
I thank Evgeny Kozik, Victor L'vov, Sergey Nazarenko, Sergey Nemirovskii, and Boris Svistunov for interesting discussions.
The work was supported by the grant of the Israel Academy of Sciences and Humanities.
\end{acknowledgments}

%

\end{document}